\documentstyle[aps,eqsecnum]{revtex}

\begin{document}
\draft
\title{Spin Relaxation Anisotropy in Two-Dimensional Semiconductor Systems}

\author{N.S.~Averkiev,$^1$ L.E.~Golub,$^{1,2}$ and M.~Willander$^2$} 
\address{$^1$A.F.~Ioffe Physico-Technical Institute, Russian Academy  of Sciences, 194021 St.~Petersburg, Russia}
\address{$^2$Physical Electronics and Photonics, Department of Physics,
Chalmers University of Technology and G\"{o}teborg University, S-412~96, G\"{o}teborg, Sweden}
\maketitle

\begin{abstract}
Spin relaxation is investigated theoretically in two-dimensional systems.
Various semiconductor structures of both n- and p-types are studied in detail.
The most important spin relaxation mechanisms are considered.
The spin relaxation times are
calculated taking into account the contributions to the spin--orbit interaction due to both the bulk
inversion asymmetry and the structure inversion asymmetry.
It is shown that in-plane anisotropy of
electron spin relaxation appears in III--V asymmetrical heterostructures.
This anisotropy may be controlled by external parameters,
{and} the spin relaxation times differ by several orders of
magnitude.
\end{abstract}

\section{Introduction}

Spin dynamics in semiconductors has been investigated for more than four decades.
During this period, effects leading to creation and disappearance of equilibrium and non-equilibrium
microscopic magnetic moments have been discovered.
The non-equilibrium spin is known to appear in III--V and II--VI compounds upon absorption
of circularly polarized light.
This way to create non-equilibrium spins is named the method of {\em Optical orientation} of
electrons and nuclei, which proved to be the most effective in investigating the optical and kinetic
properties of bulk semiconductor samples and heterostructures.
The problem of the loss of the
average microscopic spin is very important in analysis of experimental data and device applications.

In bulk semiconductors, mechanisms of electron and hole spin relaxation have been studied in
ample detail both theoretically and experimentally~\cite{Pikus/Titkov}.
It is known that the main mechanism for electrons in crystals lacking inversion symmetry is,
in a wide range of temperatures, the kinetic mechanism proposed by D'yakonov and Perel'.
Carriers lose their spin orientation owing to precession in the effective magnetic field
caused by those terms which are cubic in wavevector.
For holes in the valence band, where the spin--orbit interaction affects the
energy spectrum to a greater extent, the loss of spin is mainly associated with spin-flip in each
scattering act (Elliot--Yafet mechanism).
However, the corresponding spin relaxation rate is of the
same order as the momentum relaxation rate, which makes studying this kind of spin
relaxation complicated.

In two-dimensional (2D) semiconductor structures, the relative importance of the above
mechanisms changes dramatically because of the appearance of new forms of energy
spectrum and spin--orbit interaction.
The size-quantization effect suppresses the spin--orbit
interaction for holes and increases its strength for electrons.
Therefore, it becomes possible to study
spin relaxation in p-type structures, {and} in n-type samples it is fundamentally
different from that in the bulk case.

The spin dynamics processes in 2D systems may exhibit natural, `intrinsic' anisotropy owing to
presence of the growth axis.
Analysis shows that in the conventional heterostructures this
anisotropy is not large, being only 50\%.
However, the authors of the present study have shown
theoretically that the in-plane anisotropy of spin relaxation appears in III--V asymmetrical
heterostructures~\cite{PRB}.
This anisotropy may be controlled by external parameters, {and} the spin relaxation times
differ by several orders of magnitude~\cite{SR-FTP}.
This opens up broad possibilities for spin engineering.

We consider two most important carrier spin relaxation scenarios: the D'yakonov--Perel' and Elliot--Yafet
mechanisms.
The Bir--Aronov--Pikus mechanism, associated with exchange
electron--hole interaction, is essential only in p-doped samples, which are beyond the scope of
this review, together with the excitonic spin relaxation.

In general, spin relaxation equation can be written in the form
\begin{equation}
\dot{S}_i = - {S_j \over \tau_{ij}} \:,
\end{equation}
where $S_i$ is the spin density, and $(1 / \tau_{ij})$ is the component of a second rank tensor. The concrete form of the tensor $(1 / \tau_{ij})$ depends on the symmetry of the system under study. Since in a 2D structure the special direction exists (the growth axis), spin relaxation processes are anisotropic. Symmetry of a single heterointerface of a semiconductor with zinc-blende lattice is $C_{2v}$. This leads to three different spin relaxation times. However if two interfaces of a quantum well are equivalent, then the structure symmetry increases up to $D_{2d}$ and in-plane spin relaxes isotropically. In semiconductors, spin relaxation processes are caused firstly not by external magnetic field, but due to peculiarities of a band structure and scattering mechanisms. These are processes which are considered in the present Review.

We write the electron or hole Hamiltonian in the form
\begin{equation}
H = H_0 + V \:,
\end{equation}
where $H_0$ is the Hamiltonian of a system without weak spin--orbit interaction and scattering,
and these two are described by small perturbation $V$.
Let $| n {\bbox k} \rangle$ and $E_{n {\bf k}}$ be the eigenstates and eigenvalues of $H_0$:
\begin{equation}
\label{H0}
\langle n {\bbox k} | H_0 | n' {\bbox k}' \rangle = E_{n {\bf k}} \: \delta_{n n'} \: \delta_{\bf k k'} \:.
\end{equation}
Here $n$ is the index of a 2D electron or hole subband, and ${\bbox k}$ is the wavevector
characterizing the motion in the heterostructure plane.
In what follows, we assume that only the
first subband of size quantization is populated and the energy dispersion is isotropic.
Therefore, we omit the subband index and denote the energies as $E_{k}$.
It is convenient to extract the diagonal in the ${\bbox k}$ part of the spin--orbit interaction:
\begin{equation}
H_{\rm SO} ({\bbox k}) = V_{\bf k k} \:
\end{equation}
(small spin-independent corrections from $V_{\bf k k}$ are assumed to be included into $E_k$). 

The dynamics of the electron spin density and that of the hole angular momentum density
are determined by the time evolution of the respective density matrices.
The diagonal in the $\bbox k$ part of the density matrix, $\rho_{\bf k}$, obeys the equation

\begin{equation}
\label{rho_general}
{\partial \rho_{\bf k} \over \partial t} = {i \over \hbar } [H_{\rm SO} ({\bbox k}), \rho_{\bf k}] - {2 \pi \over \hbar} \sum_{{\bf k'} \neq {\bf k}} \left( {V_{\bf k k'} V_{\bf k' k} \rho_{\bf k} + \rho_{\bf k} V_{\bf k k'} V_{\bf k' k} \over 2} - V_{\bf k' k} \rho_{\bf k'}  V_{\bf k' k} \right) \: \delta (E_k - E_{k'}) \:.
\end{equation}
This equation is valid for any type of spin-orbit interaction and scattering. Below we analyse it in the cases of electrons and holes separately.

The review is organized as follows.
Section~II is concerned with 2D electron systems.
Section~III is devoted to p-type quantum wells.
Conclusion discusses the possible ways to measure spin
relaxation rates and to observe the spin relaxation anisotropy.

\section{Electron spin relaxation}

A specific feature of electron systems is that the spin--orbit interaction is weaker than the
momentum scattering.
Therefore, it is convenient to represent the scattering amplitude as
\begin{equation}
\label{V_amplitude}
V_{\bf k k'} = {\cal I} u_{\bf k k'} + V'_{\bf k k'} \:,
\end{equation}
where ${\cal I}$ is $2 \times 2$ unit matrix and ${\rm Tr}(V'_{\bf {k k'}}) = 0$.
Since the spin--orbit interaction is weak, $u \gg V'$.

In the following, we assume a central elastic scattering.
In this case, $u_{\bf {k k'}}$ is real and dependent on the electron energy $E_k$
and the absolute value of the scattering angle $\theta = \varphi_{\bf k'} - \varphi_{\bf k}$,
where $\varphi_{\bf k}$ is the angular co-ordinate of ${\bbox k}$ in the heterostructure plane.

Since $H_{\rm SO} ({\bbox k})$ and $V'_{\bf {k k'}}$ are small perturbations, the spin relaxation
times are much longer than the times in which the electron momentum distribution becomes
isotropic.
For this reason, it is convenient to represent the density matrix as a
sum~\cite{Pikus/Titkov}:

\[\rho = \overline{\rho} + \rho' \:, \hspace{2cm} \overline{\rho'} = 0
\:. \]
Here the bar means averaging over $\varphi_{\bf k}$, and, hence, $\overline{\rho}$ depends only on
$E_k$.
The anisotropic part of the density matrix is due to $H_{\rm SO}$ and $V'$ and, therefore,
$\rho'$ is small as compared with $\overline{\rho}$.

To second order in $H_{\rm SO}$ and $V'$, equation~(\ref{rho_general}) has the form
\begin{eqnarray}
\label{rho_el_general}
{\partial \overline{\rho} \over \partial t} + {\partial \rho'_{\bf k} \over \partial t} &=& 
{i \over \hbar } [H_{\rm SO} ({\bbox k}), \overline{\rho}] + 
{i \over \hbar } [H_{\rm SO} ({\bbox k}), \rho'_{\bf k}] \nonumber \\
&-& {2 \pi \over \hbar} g \oint \: {d \varphi_{\bf k'} \over 2 \pi} \: |u_{\bf k k'}|^2 (\rho'_{\bf k} - \rho'_{\bf k'}) \nonumber \\
&-& {2 \pi \over \hbar} g \oint \: {d \varphi_{\bf k'} \over 2 \pi} \: 
u_{\bf k k'}  \left( {V'_{\bf k k'} - V'_{\bf k' k} \over 2} \: \overline{\rho} + \overline{\rho} \: {V'_{\bf k k'} - V'_{\bf k' k} \over 2} \right)
\nonumber \\
&-& {2 \pi \over \hbar} g \oint \: {d \varphi_{\bf k'} \over 2 \pi} \: 
u_{\bf k k'}  \left( {V'_{\bf k k'} + V'_{\bf k' k} \over 2} \: \rho'_{\bf k} + \rho'_{\bf k} \: {V'_{\bf k k'} + V'_{\bf k' k} \over 2} 
- V'_{\bf k k'} \rho'_{\bf k'} - \rho'_{\bf k'} V'_{\bf k k'} \right)
\nonumber \\
&-& {2 \pi \over \hbar} g \oint \: {d \varphi_{\bf k'} \over 2 \pi} \: \left( {V'_{\bf k k'} V'_{\bf k' k} \: \overline{\rho} + \overline{\rho} \: V'_{\bf k k'} V'_{\bf k' k} \over 2} - V'_{\bf k k'} \: \overline{\rho} \: V'_{\bf k' k} \right) \:,
\end{eqnarray}
where $g$ is the 2D density of electron states.
It is noteworthy that the first two integrals in~(\ref{rho_el_general})
are of first order in spin--orbit interaction, and the two last, of second order.

The electron spin density is given by
\begin{equation}
\label{S_via_rho}
{\bbox S}_e = g \int\limits_0^\infty dE_k {\rm Tr} \left( \overline{\rho} \: {{\bbox \sigma} \over 2} \right) \:.
\end{equation}
Therefore, to analyze the spin dynamics, we have to find $\overline{\rho}$. Since $H_{\rm SO}
({\bbox k})$ and $V'_{\bf {k k'}}$ are anisotropic in ${\bbox k}$-space, the equations for
$\overline{\rho}$ and $\rho'$ are coupled and can be solved only for concrete interactions $H_{\rm
SO}$ and $V'$.

In semiconductor heterostructures, spin relaxation can be caused by both $H_{\rm SO}$ and $V'$.
The first scenario is known as the D'yakonov--Perel' mechanism, and the second, as the Elliot--Yafet
mechanism.
Even in the same heterostructure, the relative contributions of the mechanism
may vary with such parameters as temperature or electron concentration.
Since any of the spin
relaxation mechanisms may be dominant, we investigate them separately.

\subsection{D'yakonov-Perel' mechanism}

Let us consider a system with spin--orbit interaction described by the Hamiltonian $H_{\rm SO}
({\bbox k})$.
It is equivalent to a Zeeman term with effective magnetic field dependent on
${\bbox k}$.
In the presence of scattering, the wavevector changes and, hence, the effective magnetic field
changes too.
Therefore, in the case of frequent scattering, electrons are subjected to a
chaotically changing magnetic field.
The spin dynamics in such a system has diffusion character,
which leads to loss of specific spin orientation.
This is the D'yakonov--Perel' spin relaxation mechanism~\cite{DP},
and it is the main spin relaxation mechanism in many III--V bulk
semiconductors and heterostructures.
It takes place even in the case of spin-independent scattering,
when
 \[ V' = 0
\:. \]

In this case, in a time of the order of the transport time $\tau_{\rm tr}$, the density matrix becomes
isotropic, but the spin relaxation processes do not onset.
One can obtain a kinetic equation for the
spin density at times longer than $\tau_{\rm tr}$~\cite{PRB}:

\begin{equation}
\label{S}
\dot{S}_{e,i}(t) = - {1 \over 2 \hbar^2}\sum\limits_{n=-\infty}^\infty \frac{ \int\limits_0^\infty d E_k \: (F_+ - F_-) \: \tau_n \: {\rm Tr} \left( [H_{-n},[H_n, \sigma_j]] \: \sigma_i \right)}{ \int\limits_0^\infty d E_k \: (F_+ - F_-)}\: S_{e,j}(t) \:. \end{equation}

Here
\begin{equation}
\label{H'/n}
H_n = \oint \: {d \varphi_{\bf k} \over 2 \pi} \: H_{\rm SO}({\bbox k})
\exp (-i n \varphi_{\bf k} ) \end{equation}
are the Fourier-harmonics of the spin-orbit Hamiltonian, 

\begin{equation}
\label{tau/n}
{1 \over \tau_n} = {2 \pi \over \hbar} g \oint \: {d \theta \over 2 \pi} \: |u_{\bf k k'}|^2 (1 - \cos n \theta )
\end{equation}
are the transport scattering rates, and $F_\pm (E_{k})$ are the distribution functions of particles with
spin projection equal to $\pm 1/2$.

Equation~(\ref{S}) is valid for 2D electrons with any spin--orbit interaction $H_{\rm SO}({\bbox k})$.
Let us consider an asymmetrical zincblende heterostructure.
There are two contributions to $H_{\rm SO}({\bbox k})$.
The first, the so-called bulk inversion asymmetry (BIA) term, is due to
the lack of inversion symmetry in the bulk {of the} material of which the heterostructure is
made.
To calculate this term, one has to average the corresponding bulk expression over the size-quantized motion~\cite{DK}.
We investigate the heterostructure with growth direction [001]
coinciding with the $z$-axis and assume that only the first electron subband is populated.
The BIA term has the form
\begin{equation}
\label{H1}
H_{\rm BIA}({\bbox k}) = \gamma \: [ \langle k_z^2 \rangle (\sigma_y k_y - \sigma_x k_x) +  \sigma_x k_x k_y^2  - \sigma_y k_y k_x^2)] \:,
\end{equation}
where we choose the $x$- and $y$-directions aligned with the principal axes in the heterostructure
plane.
Here $\langle k_z^2 \rangle$ is squared operator $(- i \partial / \partial z)$ averaged over the
ground state, and $\gamma$ is the bulk spin--orbit interaction constant.
It is seen that $H_{\rm BIA}$ contains terms that are both linear and cubic in $k$.

In asymmetrical heterostructures, there is an additional contribution to the spin--orbit Hamiltonian,
which is absent in the bulk.
It is caused by structure inversion asymmetry (SIA) and can be written
as~\cite{OhkawaUemura,Vas'ko,BychkovRashba}
\begin{equation}
\label{H2}
H_{\rm SIA}({\bbox k})  = \alpha \: (\sigma_x k_y - \sigma_y k_x) \:,
\end{equation}
where $\alpha$ is proportional to the electric field, $E$, acting on an electron:
\begin{equation}
\label{alpha}
\alpha = \alpha_0 e  E \:.
\end{equation}
Here $e$ is the elementary charge and $\alpha_0$ is a second spin-orbit constant determined by both bulk spin-orbit interaction parameters and properties of  heterointerfaces. It should be stressed that, in asymmetrical heterostructures, $E$ is caused mainly by the difference of the wave function and band parameters at the interfaces, rather than by average electric field~\cite{Pfeffer}.

$H_{\rm SIA}$ contains terms linear in $k$, as also does $H_{\rm BIA}$.
From equation~(\ref{S}) follows that the harmonics with the same $n$
are coupled in the spin dynamics equations.
This leads to interference of BIA and SIA terms linear in wavevector in the
spin relaxation equations~\cite{PRB}.

For $H_{\rm SO} = H_{\rm BIA} + H_{\rm SIA}$, the system has $C_{2v}$-symmetry.
Therefore, equations~(\ref{S}) can be rewritten as follows:
\begin{equation}
\label{S1}
\dot{S}_z = - {S_z \over \tau_z} \:,\hspace{2cm} \dot{S}_x \pm \dot{S}_y = - {S_x \pm S_y \over \tau_\pm} \:.
\end{equation}
The times $\tau_z$, $\tau_+$ and $\tau_-$ are the relaxation times of a spin parallel to the axes
[001], [110] and [1$\bar{1}$0], respectively.

If both the spin subsystems attain equilibrium before the onset of spin relaxation, then
\begin{equation}
\label{F}
F_\pm (E_k) = F_0 (\mu_\pm -  E_k)\:,
\end{equation}
where $F_0$ is the Fermi--Dirac distribution function and $\mu_\pm$ are the chemical potentials
of the electron spin subsystems.
If the spin splitting is small, i.\,e.
$$
|\mu_+ - \mu_-| \ll |\mu_+|, |\mu_-|
$$
then the expressions for the spin relaxation rates $1/ \tau_i$ ($i = z, +, -$) have the form
\begin{equation}
\label{tau}
{1 \over \tau_i} = \frac{ \int\limits_0^\infty d E_k  \: (\partial F_0 / \partial E_k) \: \Gamma_i (k)}{ \int\limits_0^\infty d E_k \: (\partial F_0 / \partial E_k)}\:, \end{equation}
where
\begin{eqnarray}
\label{Gamma}
\Gamma_z (k) = {4 \tau_1 \over \hbar^2}\left[  (\gamma^2 \langle k_z^2 \rangle^2 +\alpha^2) \: k^2 - {1 \over 2} \gamma^2 \langle k_z^2 \rangle k^4 + {1 + \tau_3/\tau_1 \over 16} \gamma^2 k^6 \right]\:, \nonumber \\
\Gamma_+ (k) = {2 \tau_1 \over \hbar^2}\left[  (\alpha - \gamma \langle k_z^2 \rangle)^2 \: k^2 + {1 \over 2} \gamma (\alpha - \gamma \langle k_z^2 \rangle) k^4 + {1 + \tau_3/\tau_1 \over 16} \gamma^2 k^6 \right]\:,\\
\Gamma_- (k) = {2 \tau_1 \over \hbar^2}\left[  (\alpha + \gamma \langle k_z^2 \rangle)^2 \: k^2 - {1 \over 2} \gamma (\alpha + \gamma \langle k_z^2 \rangle) k^4 + {1 + \tau_3/\tau_1 \over 16} \gamma^2 k^6 \right]\:. \nonumber
\end{eqnarray}

Equations~(\ref{tau}),~(\ref{Gamma}) are valid for any electron energy distribution. If the electron
gas is degenerate, then the spin relaxation times are given by
\begin{equation}
\label{tauFermi}
{1 \over \tau_i} = \Gamma_i (k_{\rm F})\:, 
\end{equation}
where $k_{\rm F}$ is the Fermi wavevector
determined by the total 2D electron concentration $N$:
\begin{equation}
\label{kF}
k_{\rm F} = \sqrt{2 \pi N}\:.
\end{equation}
In this case, the scattering time $\tau_1$ in equations~(\ref{Gamma}) coincides with the transport
relaxation time $\tau_{\rm tr}$, which can be determined from the electron mobility.

For nondegenerate electrons, the spin relaxation times are determined, in particular, by the energy
dependences of the scattering times $\tau_1$ and $\tau_3$.
If $\tau_1$, $\tau_3\propto E_k^\nu$, then $\tau_3/\tau_1 =\rm const$ and
\begin{eqnarray}
\label{tauBoltzmann}
{1 \over \tau_z} &=& {4 \tau_{\rm tr} \over \hbar^2}\left[  (\gamma^2 \langle k_z^2 \rangle^2 +\alpha^2) \: 
{2 m k_{\rm B}T \over \hbar^2} 
- {\nu+2 \over 2} \gamma^2 \langle k_z^2 \rangle \: \left({2 m k_{\rm B}T \over \hbar^2} \right)^2 
+ (\nu+2)(\nu+3) \: {1 + \tau_3/\tau_1 \over 16} \gamma^2 \: \left({2 m k_{\rm B}T \over \hbar^2} \right)^3 \right]\:, \\
{1 \over \tau_\pm} &=& {2 \tau_{\rm tr} \over \hbar^2}\left[  (\pm \alpha - \gamma \langle k_z^2 \rangle)^2  \: 
{2 m k_{\rm B}T \over \hbar^2}
+ {\nu+2 \over 2} \gamma (\pm \alpha - \gamma \langle k_z^2 \rangle) \: \left({2 m k_{\rm B}T \over \hbar^2} \right)^2  
+  (\nu+2)(\nu+3) \: {1 + \tau_3/\tau_1 \over 16} \gamma^2 \: \left({2 m k_{\rm B}T \over \hbar^2} \right)^3 \right]\:. \nonumber
\end{eqnarray}
Here $T$ is the electron temperature and $k_{\rm B}$ is the Boltzmann constant.
In the particular case of short-range scattering, $\nu = 0$, and $\tau_1 = \tau_3$
are equal to $\tau_{\rm tr}$, which is independent of temperature.

The spin relaxation times are very sensitive to the relationship between two spin--orbit interaction
strengths, $\gamma \langle k_z^2 \rangle$ and $\alpha$.
From equations~(\ref{tauFermi}),~(\ref{tauBoltzmann}) follows that at low concentration or temperature,
$1/\tau_z$, $1/\tau_-$ and $1/\tau_+$ are determined by respectively, the sum of squared $\gamma \langle k_z^2
\rangle$ and $\alpha$, their squared sum, and their squared difference.
This may lead to a considerable difference between the three rates, i.\,e. to a total spin relaxation anisotropy if
$\gamma \langle k_z^2 \rangle$ and $\alpha$ are close in magnitude.
In real III--V systems, the relations between $H_{\rm BIA}$ and $H_{\rm SIA}$ may vary, with $H_{\rm BIA}$
or $H_{\rm SIA}$ dominant~\cite{BIAdominant,SIAdominant} or these
two terms comparable~\cite{Knap}.
In the last case, a giant spin relaxation anisotropy is predicted~\cite{PRB}.

The value of $\langle k_z^2 \rangle$ depends on the heteropotential and can be calculated for
concrete asymmetrical heterostructures~\cite{SR-FTP}.
The constant $\gamma$ is known for GaAs
from optical orientation experiments~\cite{Pikus/Titkov}.
Correct theoretical expressions for
$\gamma$ and $\alpha_0$ have been derived in terms of the three-band ${\bbox k} \cdot {\bbox
p}$ model~\cite{GS,Knap}.
The heterointerfaces make a contribution to $\alpha_0$ in addition to
that from the bulk~\cite{Silvanew}.
At large wavevectors, $\alpha_0$ starts to depend on
$k$~\cite{WinklerRoessler,Wissinger}.
Here we assume that the concentrations and temperatures
are sufficiently low, allowing us to ignore this effect.

In Ref.~\cite{SR-FTP}, the spin relaxation times were calculated for a III--V heterojunction and a
triangular quantum well (QW).
The observance of spin relaxation anisotropy in all the
three directions is predicted in a wide range of structure parameters and temperatures.


It has been shown~\cite{LMR,Silvaoldest,Jusserand} that inclusion of both the BIA and SIA
terms,~(\ref{H1}) and~(\ref{H2}), into $H_{\rm SO}$ leads to a spin-splitting anisotropy
of the conduction band in ${\bbox k}$-space in III--V semiconductor heterojunctions.
However, the spin relaxation analysis performed in~\cite{Silvaoldest} ignored this effect.
The authors of Ref.~\cite{Pikus/theor} demonstrated that the BIA and SIA terms interfere
in weak localization but are additive in spin relaxation.
It was shown in Ref.~\cite{PRB} that those
terms in $H_{\rm SO}$ which are linear in the wavevector cancel out {in} spin relaxation
as well.

In a recent study~\cite{Japan110}, the spin relaxation anisotropy was observed for uncommon
(110) GaAs QWs.
In this experiment, the spin relaxation in the growth direction was suppressed
because of the `built-in' anisotropy of the sample, resulting from the presence of heterointerfaces.
By contrast, suppression of spin relaxation in the heterostructure plane was predicted in
Ref.~\cite{PRB}.
Moreover, all the three spin relaxation times are different in the last case,
and this effect takes place in ordinary (001) heterostructures.

We demonstrate that the terms in the spin--orbit Hamiltonian, which are linear in the wavevector
interfere, and this leads to a huge anisotropy of the spin relaxation times.
At high concentration or
temperature, this effect starts to disappear owing to the predominance of the cubic in $k$ terms in
$H_{\rm SO}$, which are only present in $H_{\rm BIA}$.
However, the higher-order terms in
$H_{\rm SIA}$ are not forbidden by symmetry, either.
These terms can also interfere with those in
$H_{\rm BIA}$, and cause additional non-monotonic features in the dependences of the spin
relaxation times on the structure parameters.

\subsection{Elliot-Yafet mechanism}

Let consider a 2D electron system without spin splitting of the spectrum.
In this case, there is no $H_{\rm SO} ({\bbox k})$ term in the Hamiltonian, and the spin--orbit interaction occurs due to
scattering only.

In the heterostructures based on III--V semiconductors, the spin-flip scattering can be obtained in
the Kane model taking into account a mixture of conduction- and valence-band states.
The electron wave function has the form~\cite{ELGE}
\begin{equation}
\Psi_{s {\bf k}} ({\bbox r}) = c_k \: \exp{(i {\bbox k \cdot {\bbox \rho}})} \left[ u_s(z) S + {\bbox v}_{s {\bf k}} (z) \cdot {\bf R}\right] \:.
\end{equation}
Here $s = +,-$ enumerates two spin states at a given ${\bbox k}$; $S$ and ${\bbox R} = (X, Y, Z)$
are, respectively, $s$- and $p$-like Bloch functions, and $c_k$ is the normalization factor.
The functions $u_s$ and ${\bbox v}_{s {\bf k}}$ are eight envelopes corresponding to the conduction
and valence band, respectively, related by
\begin{equation}
{\bbox v}_{s {\bf k}} (z) = i \left( A \: {\bbox K} - i B \: {\bbox \sigma} \times {\bbox K} \right) u_s(z) \:.
\end{equation}
Here we introduce a 3D vector ${\bbox K} = (\bbox k, k_z)$ with $k_z = -i \partial /
\partial z$, and two constants:
\begin{equation}
A = {P \over 3} \left( {2 \over E_g} + {1 \over E_g + \Delta} \right) \:,
\hspace{1cm}
B = - {P \over 3} \left( {1 \over E_g} - {1 \over E_g + \Delta} \right) \:,
\end{equation}
where $E_{\rm g}$ and $\Delta$ are the energy gaps between the bands $\Gamma_6$ and $\Gamma_8$,
and $\Gamma_8$ and $\Gamma_7$, respectively, and $P$ is the Kane matrix element.
The envelope function for the conduction band is
\begin{equation}
u_s (z) = \varphi(z) \: w_s \:,
\end{equation}
where $\varphi(z)$ is the wave function of size-quantization, and $w_+ = \uparrow$, $w_- =
\downarrow$ are the spin functions.
The normalization factor is given to second order in $A K$ and $B K$ by
\begin{equation}
c_k = 1 - (A^2 + 2 B^2) (k^2 + \langle k_z^2 \rangle) \:,
\end{equation}
where $\langle k_z^2 \rangle = \int dz (d\varphi / dz)^2$.

The scattering matrix element $\langle s \bbox k | V | s' \bbox k' \rangle$ is a $2 \times 2$ matrix with respect to the indices $s$ and $s'$.
Therefore, in the notation~(\ref{V_amplitude}) we obtain
\begin{eqnarray}
\label{u-V'}
u_{\bf k k'} &=& u_0 (\bbox k - \bbox k') [1 + (A^2 + 2B^2) (\bbox k \cdot \bbox k' - k^2)] 
+ (A^2 + 2B^2) \left[ u_0 (\bbox k - \bbox k') \langle k_z^2 \rangle - \int dz u_0 (\bbox k - \bbox k',z) (k_z \varphi)^2 \right] \:, \\
V'_{\bf k k'} &=& (2AB + B^2) \left\{ \int dz u_0 (\bbox k - \bbox k',z) \: \varphi \: i k_z \varphi \: [\bbox \sigma \times (\bbox k + \bbox k')]_z  + i u_0 (\bbox k - \bbox k') \: \bbox \sigma \cdot (\bbox k \times \bbox k')
\right\} \nonumber \:.
\end{eqnarray}
Here
\begin{equation}
\label{u0}
u_0 (\bbox q, z) = \int d \bbox \rho \; V(\bbox \rho, z)\exp{(i \bbox q \cdot \bbox \rho)} \:,
\end{equation}
with $V(\bbox \rho, z)$ being the scattering potential, and
\begin{equation}
u_0 (\bbox q) = \int d z \: u_0(\bbox q, z) \: \varphi^2 (z)  \:.
\end{equation}
The difference between $u_{\bf {k k'}}$ and $u_0 (\bbox k - \bbox k')$ is due to the
spin--orbit interaction, similarly to the 3D case~\cite{A&Ya}.

The two terms in $V'$ are different: the part containing $\sigma_x$ and $\sigma_y$ is linear in $k$,
but the term with $\sigma_z$ is $\sim k^2$.
Hence, at $k^2 \ll \langle k_z^2 \rangle$, the first part
is dominant, and $V'_{\bf {k k'}}$ can be represented as
\begin{equation}
\label{V'-v}
V'_{\bf k k'} =  v(\bbox k - \bbox k') \: [\bbox \sigma \times (\bbox k + \bbox k')]_z  \:,
\end{equation}
where
\begin{equation}
\label{v}
v(\bbox q) = (2AB + B^2) \int dz u_0 (\bbox q , z) \: \varphi \: i k_z \varphi \:.
\end{equation}

It is noteworthy that a nonzero contribution at $\bbox k = \bbox k'$ is the impurity-induced SIA
term physically meaning the force exerted on electrons by scatterers.
This term is to be added to
$H_{\rm SO}$, and it manifests itself in the D'yakonov--Perel' mechanism only.

It is seen from equation~(\ref{V'-v}) that
\[
V'_{\bf k k'} = V'_{\bf k' k}
\]
for central scattering, whereas $v(\bbox k - \bbox k') = v(E_k, | \varphi_{\bf k} - \varphi_{\bf k}'|)$. Therefore, averaging of Eq.~(\ref{rho_el_general}) over $\varphi_{\bf k}$ gives
\begin{eqnarray}
\label{rho_EY}
{\partial \overline{\rho} \over \partial t} = 
- {2 \pi \over \hbar} g \oint \: {d \varphi_{\bf k} \over 2 \pi} \:\oint \: {d \varphi_{\bf k'} \over 2 \pi} &\biggl\{& {V^{'2}_{\bf k k'} \: \overline{\rho} + \overline{\rho} \: V^{'2}_{\bf k k'} \over 2} - V'_{\bf k k'} \: \overline{\rho} \: V'_{\bf k k'} 
- u_{\bf k k'}  \left[ (\rho'_{\bf k'} - \rho'_{\bf k}) V'_{\bf k k'} + V'_{\bf k k'} (\rho'_{\bf k'} - \rho'_{\bf k}) \right] \biggl\} \:.
\end{eqnarray}
It can be shown that the term in square brackets is proportional to the integral
\[
\oint \: d \theta u(\theta) v(\theta) \sin{\theta}
\]
which is zero for central scattering. 

Using Eq.~(\ref{S_via_rho}), we obtain
\begin{equation}
{1 \over \tau_{zz}} = {2 \over \tau_{xx}} = {2 \over \tau_{yy}} =
{16 \pi m \over \hbar^3}
{\int\limits_0^\infty dE_k \: E_k \: [F_+(E_k) - F_-(E_k)] \oint \: d \theta \: v^2(E_k, \theta) (1 + \cos{\theta})
\over
\int\limits_0^\infty dE_k \: [F_+(E_k) - F_-(E_k)]} \:.
\end{equation}
It can be seen that the Elliot--Yafet mechanism results in a 50~\% anisotropy of spin
relaxation times.

To second order in $\Delta / E_{\rm g}$, we have for a Boltzmann gas:
\begin{equation}
{1 \over \tau_{zz}} = {32 \over 9} \left( \Delta \over E_g \right)^2 {k_{\rm B} T \over E_g} \: r \: {1 \over \tau_{\rm tr}} \:,
\end{equation}
with
\begin{equation}
r = {\hbar^2 \over m E_g}
{\int\limits_0^\infty dE_k \: E_k \: \exp{(- E_k / k_{\rm B} T)} \oint \: d \theta \:(1 + \cos{\theta}) \left[  \int dz u_0(q,z) \varphi(z) ik_z \varphi(z) \right]^2
\over
\int\limits_0^\infty dE_k \: E_k \: \exp{(- E_k / k_{\rm B} T)} \oint \: d \theta \:(1 - \cos{\theta}) \left[  \int dz u_0(q,z) \varphi^2(z)\right]^2
}
\:.
\end{equation}
Here the wavevector transfer in the case of elastic scattering is described by
\[
q = \sqrt{8mE_k \over \hbar^2} \left| \sin{\theta \over 2}\right| \:,
\]
and account was taken of the fact that, at a given accuracy,
\[
{\hbar^2 \over m} = {P^2 \over E_g} \:.
\]
The factor $r$ is on the order of $E_1 / E_{\rm g}$, where $E_1$ is the energy
of the first level of size-quantization.
For an infinitely deep rectangular QW of width 100~\AA, $E_{\rm g} = 1.5$~eV, and
for randomly distributed $\delta$-scatterers, $r=1.2 \times 10^{-2}$.
Hence, for the parameters of GaAs ($\Delta = 0.34$~eV, $m=0.067 m_0$) and $T=77$~K,
we have $\tau_{\rm tr} / \tau_{zz} \sim 10^{-5}$.

In addition to the scattering processes leading to spin relaxation considered above, there is a short-range interaction making possible scattering of electrons from the $\Gamma_8$ or $\Gamma_6$ band to $\Gamma_7$ band~\cite{Pikus/Titkov}. The same mechanism takes place in the 2D systems as well~\cite{IT}.

\section{Hole spin relaxation}

A specific feature of p-type QWs is a strong spin--orbit interaction: the states at the top of the
valence size-quantized subbands have a certain projection of the angular momentum on
the growth axis.
Therefore, ordinary spin-independent scattering leads to hole angular momentum
relaxation even in a symmetrical QW, when there are no additional spin--orbit terms: $H_{\rm SO}= 0$.
In other words, the Elliot--Yafet mechanism is very important for holes.

In a symmetrical $p$-type QW, each hole level is doubly degenerate.
The wave functions have the form
\begin{equation}
\Psi_{s {\bf k}} = \exp{(i \bbox k \cdot \bbox \rho)} \: F_s (\bbox k, z) \:,
\end{equation}
where $s=1,2$ enumerates the degenerate states, and $Oz$ is the growth direction.

In a QW based on a III--V semiconductor, the hole wave function is a superposition of
those four states of the valence band top, which correspond to the angular momentum projections
on $Oz$ equal to $3/2$, $1/2$, $-1/2$ and $-3/2$.
In the basis of these states, $F_s ({\bbox k}, z)$ are
four-component columns which can be represented as~\cite{MPP}

\begin{equation}
\label{F_12}
F_1 =\left[ \begin{array}{c} - v_0 \: C(z) \\ i v_1 \: S(z) \:
e^{i\varphi_k} \\ - v_2 \: C(z) \:e^{2i\varphi_k} \\ i v_3 \: S(z) \:
e^{3i\varphi_k} \end{array} \right]\:,
\:\:\:\:
F_2 = \left[ \begin{array}{c} i v_3 \: S(z) \: e^{- 3 i
\varphi_k} \\ v_2 \: C(z) \:e^{- 2 i \varphi_k} \\ i v_1 \: S(z) \:
e^{- i \varphi_k} \\ v_0 \: C(z) \end{array} \right]\:.
\end{equation}
Here $C(z)$ and $S(z)$ are, respectively, symmetrical and asymmetrical functions of the coordinate
$z$, and the real coefficients $v_m$ ($m=0 \dots 3$) for the ground subband are given by
\begin{equation}
\label{low_k}
v_m \sim k^m, \hspace{1cm} k \to 0 \:.
\end{equation}

An analog of the electron spin density in experiments on circular polarization
of the luminescence in low magnetic fields is the pseudospin of holes, $\bbox S_h$.
Its value is proportional to the degree of circular polarization of radiation {at relatively weak} hole
orientation.
The  {hole pseudospin} density is given by
\begin{equation}
\label{S_h}
\bbox S_h = \sum_{\bf k} {\rm Tr} \: [\rho_{\bf k} \: \bbox J (\bbox k)] \:,
\end{equation}
where $\bbox J (\bbox k)$ is the pseudospin operator for the first hole level of size-quantization. In the basis~(\ref{F_12}), $J^i (\bbox k)$ has a form of $2 \times 2$ matrix with the elements
\begin{equation}
J^i_{ss'} (\bbox k) = \int dz \: F_{s}^{\dag} (\bbox k, z) \: {\cal J}^i \: F_{s'} (\bbox k, z) \:,
\end{equation}
where ${\cal J}^i$ are $4 \times 4$ matrices in the basis of {the} four states of the valence
band top~\cite{ELGE}.
Calculation yields

\begin{equation}
\label{J_z}
J^z (E_k) = \sigma_z \: j_z\:,
\end{equation}
\begin{equation}
\label{J_+}
{J^x + i J^y \over 2} =
\left( 
\begin{array}{cc} 
0 & a e^{-2i\varphi_k} \\ 
b e^{4i\varphi_k} &  0
\end{array} 
\right)\:,
\end{equation}
\begin{equation}
\label{J_-}
{J^x - i J^y \over 2} =
\left( 
\begin{array}{cc} 
0 & b e^{-4i\varphi_k} \\ 
a e^{2i\varphi_k} &  0
\end{array} 
\right)\:,
\end{equation}
where
\begin{equation}
j_z(E_k) =  \left( {3 \over 2} v_0^2 - {1 \over 2} v_2^2 \right) \int dz C^2(z) + \left( {1 \over 2} v_1^2 - {3 \over 2} v_3^2 \right) \int dz S^2(z) \:,
\end{equation}
\begin{equation}
a (E_k) = v_1^2 \int dz S^2(z) - \sqrt{3} v_0 v_2 \int dz C^2(z) \:,
\end{equation}
\begin{equation}
b (E_k) = \sqrt{3} v_1 v_3 \int dz S^2(z) - v_2^2 \int dz C^2(z) \:.
\end{equation}
It can be seen that $J^z$ depends on the energy $E_k$, whereas $J^x$ and $J^y$ depend on both
$E_k$ and the direction of $\bbox k$.
Therefore, we have to find not only the isotropic part of the
density matrix, $\overline{\rho}$, but also the 2nd and 4th Fourier-harmonics.

$J^i$ have complicated energy dependences.
For instance, $j_z(E_k = 0) = 3/2$ for the first heavy-hole
subband of size-quantization.
The dependence $j_z$ on $k$ is given in~\cite{F&B} for
different size-quantization subbands.
It follows from equation~(\ref{low_k}) that $a \sim k^2$ and $b
\sim k^4$ for small $k$, i.\,e. $J^x$, $J^y$ are nonzero owing to nonparabolicity effects only.

\subsection{Elliot-Yafet mechanism}

Let us consider scattering from randomly distributed short-range impurities
\begin{equation}
\label{V_short_range}
V(\bbox r) = \sum_i V_0 \: \delta(\bbox r - \bbox R_i) \:.
\end{equation}
Calculating the scattering amplitude $V_{ss'} (\bbox k, \bbox k')$ and using Eq.~(\ref{rho_general}), we obtain the following equation for the hole density matrix

\begin{equation}
 \label{rho_h}
{\partial \rho_{\bf k} \over \partial t} = - {2 \pi \over \hbar} g(E_k) \oint \: {d \varphi_{\bf k'} \over 2 \pi} \: \left( \langle |V_{11}|^2 + |V_{12}|^2 \rangle \rho_{\bf k} - \langle V_{\bf k k'} \rho_{\bf k'} V_{\bf k' k} \rangle \right) \:.
\end{equation}
Here, the angular brackets denote averaging over $\bbox R_i$, and account is taken of the fact that, for~(\ref{F_12}) and~(\ref{V_short_range}),
\[
\langle V_{11} \cdot V_{12} \rangle = \langle V_{22} \cdot V_{12} \rangle = 0
\]
and $|V_{ss'}|$ depends on $\theta = \varphi_{\bf k} - \varphi_{\bf k'}$.
Note that the spin--orbit interaction in $p$-type QWs is strong and, therefore,
the division~(\ref{V_amplitude}) makes no sense, and we use the total scattering amplitude
$V_{\bf {kk'}}$.

We expand the density matrix in a series
\begin{equation}
\label{expand}
\rho_{\bf k} = \sum_{n=-\infty}^{\infty} \left[ {\cal I} f_n(E_k) + \bbox \sigma \cdot \bbox \ae_n (E_k) \right] \exp{(i n \varphi_{\bf k})} \:.
\end{equation}
The spin-independent part obeys the equation
\begin{equation}
\label{f_n_h}
{\partial f_n \over \partial t} = - {2 \pi \over \hbar} g(E_k) \oint \: {d \theta \over 2 \pi} \: \langle |V_{11}|^2 + |V_{12}|^2 \rangle (1 - \cos{n\theta}) \: f_n \:.
\end{equation}
It is seen that the total number of particles with a given energy $f_0(E_k)$ is conserved, which is
correct for the elastic scattering under study.
The Fourier-harmonics $f_n$ with $n \geq 1$ relax
with the corresponding transport times.

For the $z$-component of ${\bbox \ae}_n$ we have
\begin{equation}
{\partial \ae_n^z \over \partial t} = - {2 \pi \over \hbar} g(E_k) \oint \: {d \theta \over 2 \pi} \: \left[ \langle |V_{11}|^2 \rangle (1 - \cos{n\theta})  + \langle |V_{12}|^2 \rangle (1 + \cos{n\theta}) \right] \: \ae_n^z \:.
\end{equation}
It is important that for small $k$, 
\[
\Re{V_{11}} \gg \Im{V_{11}} \gg |V_{12}| \:.
\] 
Therefore, the first term in square brackets is close to the inverse transport time, and the second is
much smaller and describes spin relaxation.
It is seen that the average pseudo-spin with energy
$E_k$, $\ae_0^z$, relaxes with time much longer that $\tau_{\rm tr}$.
However, higher Fourier-harmonics change much faster.

The same is true for $\ae^{x,y}$.
This can be readily shown for small $k$, when we can neglect
$V_{12}$ compared with $V_{11}$.
In this limit
\begin{equation}
{\partial (\ae_n^x \pm i \ae_n^y ) \over \partial t} = - \Gamma_n^\pm \: (\ae_n^x \pm i \ae_n^y) \:,
\end{equation}
\begin{equation}
\Gamma_n^\pm = {2 \pi \over \hbar} g(E_k) \oint \: {d \theta \over 2 \pi} \: \left[ \langle |V_{11}|^2 \rangle (1 - \cos{n\theta})  + 2 \langle (\Im{V_{11}})^2 \rangle \cos{n\theta} \mp 2 \langle \Re{V_{11}} \cdot \Im{V_{11}} \rangle \sin{n\theta} \right] \:.
\end{equation}
This means that all harmonics relax with time $\sim \tau_{\rm tr}$ except that with $n=0$.
The zero harmonic determines the hole Cooperon, important in the weak localization effect.
Its relaxation is caused by the small term $|V_{12}|$~\cite{JETP,FTP}.
However, the in-plane pseudospin components $J^x$, $J^y$ are determined by the non-zero harmonics $\ae^{x,y}_{2,4}$
[see equations~(\ref{J_+}),~(\ref{J_-})] and, hence, their relaxation is fast.
Therefore, one can investigate only the dynamics of $S_{h,z}$ in optical experiments.
The corresponding equation is
\begin{equation}
\label{S_h_z}
{\partial S_{h,z} \over \partial t} = - {4 \pi \over \hbar}
{\int\limits_0^\infty dE_k  \: g^2(E_k) \: [F_+(E_k) - F_-(E_k)] \: j_z(E_k) \oint \: {d \theta \over 2 \pi} \: 
\langle |V_{12}|^2 \rangle 
\over
\int\limits_0^\infty dE_k \: g(E_k) \: [F_+(E_k) - F_-(E_k)] \: j_z(E_k) } \: S_{h,z} \:.
\end{equation}
In Ref.~\cite{F&B}, the value $(2 \pi /\hbar)  g(E_k)\oint {\rm d \theta/2 \pi}
\langle |V_{12}|^2 \rangle$ was calculated as a function of $k$ for different scattering potentials.

\subsection{D'yakonov-Perel' mechanism}

If we take into account the splitting between the pseudospin states in the valence subbands, which
may be caused by both BIA and SIA (in asymmetrical heterostructures), then the operator $H_{\rm
SO} \neq 0$ for holes.
For the ground subband, in the basis of the two degenerate states,
it is a $2 \times 2$ matrix.
Therefore, the problem is equivalent to that for electrons, and we can
obtain the equation for the density matrix dynamics.
Expanding the density matrix in Fourier
series~(\ref{expand}), we obtain expressions for $f_n$ and $\bbox \ae_n$.
The former coincides with
equation~(\ref{f_n_h}), and the latter is as follows

\begin{equation}
{\partial \ae_n^i \over \partial t} = - {2 \pi \over \hbar} g(E_k) \oint \: {d \theta \over 2 \pi} \: \tau_n {\rm Tr}( [H_{-n}, [H_n, \sigma_i] ] \sigma_i ) \: \ae_n^i \:.
\end{equation}
According to equations~(\ref{J_z})-(\ref{J_-}), in order to investigate the hole pseudospin dynamics, we
have to average over $E_k$ the equations with $n = 0, \pm 2$ and $\pm 4$.
It can be seen that, similarly to the Elliot--Yafet mechanism, relaxation of $S_{h,x}$ and $S_{h,y}$
is fast, taking a time $\sim \tau_{\rm tr}$.
This is due to the anisotropic nature of the operators $J^x$ and $J^y$.
The only slowly relaxing component is $S_{h,z}$.
The corresponding equation has the form
\begin{equation}
\label{S_hz}
{\partial S_{h,z} \over \partial t} = - {1 \over 2 \hbar^2} \sum_{n = - \infty}^\infty
{\int\limits_0^\infty dE_k  \: g(E_k) \: [F_+(E_k) - F_-(E_k)] \: j_z(E_k) \: \tau_n {\rm Tr}( [H_{-n}, [H_n, \sigma_z] ] \sigma_z )
\over
\int\limits_0^\infty dE_k \: g(E_k) \: [F_+(E_k) - F_-(E_k)] \: j_z(E_k)} \: S_{h,z} \:.
\end{equation}

In \cite{F&B-DP}, correct equations were obtained for relaxation of the zero harmonics of the
density matrix (for $f_0$ and ${\bbox \ae}_0$).
Therefore, the slow time calculated there describes
the relaxation of $S_{h,z}$ only.
The relaxation time of $\ae_0^{x,y}$ was incorrectly associated
with that of $S_{h,x}$ and $S_{h,y}$.

To calculate the hole pseudospin relaxation time, a knowledge is necessary of the concrete form of
$H_{\rm SO}$ for a structure studied.
The BIA contributions can be obtained by averaging the
corresponding bulk terms~\cite{ELGE}.
Moreover, the SIA contributions exist in asymmetrical heterostructures~\cite{Winkler}.
All of them lead to hole pseudospin relaxation in
accordance with equation~(\ref{S_hz}).

\section{Conclusion}
The predicted spin relaxation anisotropy can be observed in time-resolved
measurements similar to those in Ref.~\cite{Japan110}.
In steady-state experiments, the
in-plane electron spin relaxation anisotropy can be investigated by means of the Hanle
effect.
To obtain the spin relaxation times, account should be taken of the fact that the Land\'{e}
$g$-factor has not only diagonal in-plane components ($g_{xx}$), but also off-diagonal ones
($g_{xy}$)~\cite{KK}.
The degree of photoluminescence polarization in a magnetic field
${\bbox B} \perp z$ is described by the following expression
\begin{equation}
\label{Hanle}
P ({\bbox B}) = \frac{P(0)}{1 + [\mu_B \: (g_{xx} \pm g_{xy}) \: B/\hbar]^2 \: \tau_z \: \tau_\mp} \:,
\end{equation}
where the upper and lower signs correspond, respectively, to the experimental arrangements with ${\bbox B} || [110]$ and ${\bbox B} || [1\bar{1}0]$ ($\mu_B$ is the Bohr magneton). By changing the structure asymmetry (e.\,g., by applying a gate voltage), one can modify the Hanle curves owing
to the in-plane spin relaxation anisotropy.

\section*{Acknowledgements} 

We thank V.~V.~Bel'kov for fruitful discussions.
This study was supported by the RFBR (projects 00-02-17011, 00-02-16894, 01-02-17528 and 01-02-06354), by the Programms of Russian Ministry of Industry, Science and Technology, by the Programme of Presidium of RAS ``Quantum low-dimensional structures'', and by INTAS.


\newpage
\begin{references}
\bibitem[]{e-mail}E-mail: golub@coherent.ioffe.rssi.ru

\bibitem{Pikus/Titkov} G.E.~Pikus and A.N.~Titkov in {\em Optical orientation}, edited by F.~Meier, B.P.~Zakharchenya, North-Holland, Amsterdam, 1984.

\bibitem{PRB}N.S.~Averkiev and L.E.~Golub, \prb {\bf 60}, 15582 (1999).

\bibitem{SR-FTP}N.\,S.~Averkiev, L.\,E.~Golub and M.~Willander,  Fiz. Techn. Poluprov. {\bf 36}, 97 (2002) [Semiconductors {\bf 36}, 91 (2002)]. 

\bibitem{DP} M.I.~D'yakonov and V.I.~Perel', Fiz. Tverd. Tela {\bf 13}, 3581 (1971) [Sov. Phys. Solid State {\bf 13}, 3023 (1972)]; Zh. Eksp. Teor. Fiz. {\bf 60}, 1954 (1971) [Sov. Phys. JETP {\bf 33}, 1053 (1971)].

\bibitem{DK} M.I.~D'yakonov and V.Yu.~Kachorovskii, Fiz. Techn. Poluprov.
{\bf 20}, 178 (1986) [Sov. Phys. Semicond. {\bf 20}, 110 (1986)].

\bibitem{OhkawaUemura} F.J.~Ohkawa and Y.~Uemura, J. Phys. Soc. Jpn. {\bf 37}, 1325 (1974).

\bibitem{Vas'ko} F.T.~Vas'ko, Pis'ma Zh. Eksp. Teor. Fiz. {\bf 30}, 574 (1979)
[JETP Lett. {\bf 30}, 541 (1979)].

\bibitem{BychkovRashba} Yu.L.~Bychkov and \'{E}.I.~Rashba,  Pis'ma Zh. Eksp. Teor. Fiz. {\bf 39}, 66 (1984) [JETP Lett. {\bf 39}, 78 (1984)].

\bibitem{Pfeffer}P.~Pfeffer, \prb {\bf 59}, 15902 (1999).

\bibitem{BIAdominant}
B. Jusserand, D. Richards, H.~Peric, and B. Etienne,  \prl {\bf 69}, 848 (1992).

\bibitem{SIAdominant} D.~Stein, K.~von Klitzing, and G.~Weimann,  \prl {\bf 51}, 130 (1983).

\bibitem{Knap} W.~Knap, C.~Skierbiszewski, A.~Zduniak, E.~Litwin-Staszevska, D.~Bertho, F.~Kobbi, J.L.~Robert, G.E.~Pikus, F.G.~Pikus, S.V.~Iordanskii, V.~Mosser, K.~Zekentes, and Yu.B.~Lyanda-Geller, \prb {\bf 53}, 3912 (1996).

\bibitem{GS} L.G.~Gerchikov and A.V.~Subashiev, Fiz. Techn. Poluprov.
{\bf 26}, 131 (1992) [Sov. Phys. Semicond. {\bf 26}, 73 (1992)].

\bibitem{Silvanew} E. A. de Andrada e Silva, G.C.~La Rocca, and F.~Bassani,  \prb {\bf 55}, 16293 (1997).

\bibitem{WinklerRoessler}R.~Winkler and U.~R\"ossler,  \prb {\bf 48}, 8918 (1993).

\bibitem{Wissinger}L.~Wissinger, U.~R\"ossler, R.~Winkler, B.~Jusserand, and D.~Richards,  \prb {\bf 58}, 15375 (1998).

\bibitem{LMR}G.~Lommer, F.~Malcher, and U.~R\"ossler,  \prl {\bf 60}, 728 (1988).

\bibitem{Silvaoldest} E. A. de Andrada e Silva,  \prb {\bf 46}, 1921 (1992).

\bibitem{Jusserand} 
B. Jusserand, D. Richards, G. Allan, C. Priester and B. Etienne,  \prb {\bf 51}, 4707 (1995).

\bibitem{Pikus/theor} F.G.~Pikus and G.E.~Pikus,  \prb {\bf 51}, 16928 (1995).

\bibitem{Japan110} Y.~Ohno, R.~Terauchi, T.~Adachi, F.~Matsukura, and H.~Ohno,  \prl {\bf 83}, 4196 (1999).

\bibitem{ELGE} E.L. Ivchenko and G.E. Pikus, {\em Superlattices
and Other Heterostructures. Symmetry and Optical Phenomena},
Springer Series in Solid State Sciences, Vol. 110, Springer-Verlag,
Heidelberg, 1995; 2nd ed., 1997.

\bibitem{A&Ya} V.N.~Abakumov and I.N.~Yassievich, Zh. Eksp. Teor. Fiz. {\bf 61}, 2571 (1971) [Sov. Phys. JETP {\bf 34}, 1375 (1972)].

\bibitem{IT} E.L. Ivchenko and S.A. Tarasenko, to be published

\bibitem{MPP}I.A.~Merkulov, V.I.~Perel, and M.E.~Portnoi, Zh. Exp. Teor. Fiz. {\bf 99}, 1202 (1991) [Sov. Phys. JETP {\bf 55}, 669 (1991)]. 

\bibitem{JETP}N.S.~Avierkiev, L.E.~Golub, and G.E.~Pikus, Zh. Exp. Teor. Fiz. {\bf 113}, 1429 (1998) [JETP {\bf 86}, 780 (1998)]. 

\bibitem{FTP} N.\,S.~Averkiev, L.\,E.~Golub and G.\,E.~Pikus, Fiz. Techn. Poluprov. {\bf 32}, 1219 (1998) [Semiconductors {\bf 32}, 1087 (1998)]. 

\bibitem{F&B}R.~Ferreira and G.~Bastard,  \prb {\bf 43}, 9687 (1991).

\bibitem{F&B-DP}R.~Ferreira and G.~Bastard,  Europhys. Lett. {\bf 23}, 439 (1993).
\bibitem{Winkler}R.~Winkler, \prb {\bf 62}, 4245 (2000).
\bibitem{KK} V.K.~Kalevich and V.L.~Korenev, Pis'ma Zh. Eksp. Teor. Fiz. {\bf 57}, 557 (1993)
[JETP Lett. {\bf 57}, 571 (1993)].

\end{references}
\end{document}